\documentclass[superscriptaddress, preprintnumbers,amsmath,amssymb, aps, prl, floatfix, twocolumn, nofootinbib]{revtex4-2}
\usepackage{mathrsfs}
\usepackage{graphicx}
\usepackage{dcolumn}
\usepackage{bm}
\usepackage{enumitem}

\usepackage{comment}
\usepackage{braket}
\usepackage{amsthm}
\usepackage{tikz}
\usepackage{color,soul}
\usetikzlibrary{positioning}
\usepackage{hyperref}
\hypersetup{
    colorlinks=false,
    linkcolor=blue,
    filecolor=magenta,      
    urlcolor=cyan,
}

\begin{document}

\preprint{YITP-23-28}

\title{Higgs-confinement continuity and matching of Aharonov-Bohm phases
}

\author{Yui Hayashi}
\email{yui.hayashi@yukawa.kyoto-u.ac.jp}\affiliation{
Yukawa Institute for Theoretical Physics, Kyoto University, Kitashirakawa Oiwakecho, Sakyo Ward, Kyoto, 606-8502, Japan.
}

\begin{abstract}
Some gauge theories with a spontaneously broken $U(1)$ symmetry exhibit fractional Aharonov-Bohm (AB) phases around vortices in the Higgs regime.
We discuss continuity between confining and Higgs regimes in such gauge theories with fundamental matter fields, focusing on the AB phases.
By explicit calculations in relevant lattice models, we demonstrate that the AB phase is smoothly connected between the confining and Higgs regimes, supporting the Higgs-confinement continuity.
This result provides new insight into phase structures of gauge theories with superfluidity, such as dense QCD.

%

\end{abstract}

\maketitle

\section{INTRODUCTION}

Gauge theories with fundamental matters are believed to exhibit a smooth connection between the confining and Higgs regimes, referred to as Higgs-confinement continuity \cite{Fradkin:1978dva, Banks:1979fi, Osterwalder:1977pca}. The Higgs ``condensation'' cannot be used as an order parameter due to its lack of gauge-invariance \cite{Elitzur:1975im}. Unless global symmetries differ, these regimes have identical qualitative features, such as spectra of local gauge-invariant operators, perimeter-law of large Wilson loops, etc. Indeed, analyticity is rigorously proved in a region connecting the confining and Higgs regimes in the lattice gauge model with a single gauge-group-valued fundamental Higgs field, known as the Fradkin-Shenker-Osterwalder-Seiler theorem \cite{Osterwalder:1977pca, Fradkin:1978dva}.

This Higgs-confinement continuity is not only an important topic in field theory for understanding confinement and the Brout-Englert-Higgs mechanism, but it also plays a pivotal role in the quark-hadron continuity problem \cite{Schafer:1998ef, Schafer:1999fe, Alford:1999pa, Casalbuoni:1999wu} in dense quantum chromodynamics (QCD) \cite{Alford:2007xm, Fukushima:2010bq}. This problem is of phenomenological interest because the existence or absence of a phase transition between the hadronic and quark-matter phases would strongly affect the internal structure of neutron stars \cite{Baym:2017whm}. Quarks are expected to form Cooper pairs in the high-density region, analogous to a usual superconductor. This phase is characterized by diquark condensation and is called the color superconducting phase \cite{Barrois:1977xd, Alford:1998mk}. Near the $SU(3)$ flavor symmetric limit, the color-flavor locked (CFL) phase is expected to appear as a color superconducting phase.
Because the CFL and nuclear superfluid phases share the same symmetry-breaking pattern, it has been conjectured that these phases are smoothly connected, referred to as the quark-hadron continuity \cite{Schafer:1998ef}.
The Higgs-confinement continuity suggests that this conjecture is a consistent scenario since the diquark condensate characterizing the CFL phase is in color (anti-)fundamental representation.

The Higgs-confinement continuity has regained interest \cite{Cherman:2018jir, Cherman:2020hbe, Hirono:2018fjr, Hirono:2019oup, Hidaka:2022blq, Verresen:2022mcr, Greensite:2020nhg, Greensite:2018mhh, Greensite:2017ajx, Harlow:2018tng} due to its formal and phenomenological significance as well as recent progress in understanding symmetry \cite{Gaiotto:2014kfa}.
In particular, motivated by the topological ordering \cite{Wen:1989iv}, Refs.~\cite{Cherman:2018jir, Cherman:2020hbe} focus on particle-vortex braiding phases, or the AB phases, to distinguish hadronic (confining) and CFL (Higgs) phases\footnote{The idea that the AB phase (or particle-vortex statistics), including the non-Abelian case, can be an order parameter had been discussed in 
\cite{Krauss:1988zc, Alford:1990mk, Preskill:1990bm, Alford:1990ur, Alford:1992yx, Alford:1990fc}, even before the understanding of topological ordering became widespread.}.
Superfluid vortices exist in both phases due to the common breaking pattern of the quark number symmetry, $U(1) \rightarrow \mathbb{Z}_6$. 
In the hadronic phase, this vortex is the standard dibaryon superfluid vortex.
In the CFL phase, such a minimal vortex has been studied as the non-Abelian CFL vortex \cite{Balachandran:2005ev,Nakano:2007dr, Eto:2009bh,Eto:2009kg,Eto:2009tr,Cipriani:2012hr, Alford:2016dco,Chatterjee:2015lbf,Alford:2018mqj,Chatterjee:2018nxe, Eto:2013hoa}.
It has been found that the color AB phase around the vortex is nontrivial $\mathbb{Z}_3$ phase in the CFL phase \cite{Cherman:2018jir}.


The question of whether this nontrivial braiding phase implies a Higgs-confinement phase transition has received recent attention. 
Although the nontrivial AB phase in the CFL phase does not imply topological order \cite{Hirono:2018fjr, Hirono:2019oup}, it is still unclear how AB phases match between the confining and Higgs regimes. The AB phase seems trivial in the confining regime, at least from a classical vortex configuration, which suggests a discrepancy in AB phases between these regimes {\cite{Cherman:2020hbe}. 
The Fradkin-Shenker-Osterwalder-Seiler theorem does not apply here, as the Higgs fields are charged under a spontaneously broken global symmetry. 
Thus, it is worthwhile to extend the Fradkin-Shenker argument for the AB phase in such gauge-Higgs systems to answer the above question.}

This paper addresses the continuity problem in lattice gauge-Higgs models which have nontrivial AB phases in the Higgs regime.
These models are based on a toy model proposed in Ref.~{\cite{Cherman:2020hbe}} and an effective model of the CFL phase.
In these models, we demonstrate the continuity of the AB phase by explicit calculations in strong coupling and deep Higgs regions, which connect the confining and Higgs regimes.
This result offers new insights into topological nature of the Higgs-confinement continuity and suggests that the quark-hadron continuity remains a viable possibility.

\section{Aharonov-Bohm phase around superfluid vortex}

In this section, we define the AB phase around a vortex and introduce the main results.
For a setup, let us consider models with the following properties:
\begin{enumerate}
    \item This model is a gauge theory with at least one fundamental representation matter. 
    \item This model has a global $U(1)_G$ symmetry nontrivially acting on the fundamental matter, and there is a phase with spontaneously broken $U(1)_G$ symmetry\footnote{Due to this assumption, the spacetime dimension $d$ must be larger than two \cite{Coleman:1973ci, Mermin:1966fe}: $d > 2$.}.
    \item Within this $U(1)_G$-broken phase, it is possible to tune a parameter so that the fundamental representation matter decouples/higgses without changing the realization of any global symmetry.
\end{enumerate}

A superfluid vortex exists in the $U(1)_G$-broken phase.
Let us define $V(S)$ as the minimal vortex operator, where $S$ is the vortex worldsheet.
Note that, if the $U(1)_G$ symmetry is broken to its discrete subgroup $\mathbb{Z}_q$, the ``minimal vortex'' refers to a unit winding of $\pi_1(U(1)_G/\mathbb{Z}_q) = \mathbb{Z}$.
For instance, the minimal vortex in the CFL phase is the non-Abelian CFL vortex.

We define the AB phase $O_\Omega$ as a normalized Wilson loop $W(C)$ with a fixed winding number \cite{Cherman:2020hbe}.
In the low-energy effective theory, it is the particle-vortex statistical phase:
\begin{align}
    O_\Omega = \frac{\braket{W(C)V(S)}}{\braket{W(C)} \braket{V(S)}}, \label{eq:def_order
_param}
\end{align}
where the Wilson loop has a unit winding with the vortex: $\operatorname{Link}(C,S) = 1$.
The denominator is introduced for the normalization $|O_\Omega| = 1$.
This quantity measures the AB phase of a fundamentally charged particle surrounding the vortex.
To measure $O_\Omega$ in the original ultraviolet theory, we shall compute
\begin{align}
    O_\Omega = \lim_{|C| \rightarrow \infty} \frac{\braket{W(C)}_{w(C) = 1}}{|\braket{W(C)}_{w(C) = 1}|}, \label{eq:UV_def}
\end{align}
where $|C| \rightarrow \infty$ stands for taking an asymptotically large closed loop $|C|$, and $\braket{\cdot}_{w(C) = 1}$ is the expectation value over configurations with the fixed winding number\footnote{Since the superfluidity is assumed, the winding number should be well-defined, at least for large $C$.} $w (C) = 1$.

We consider theories in which the AB phase is nontrivial $O_\Omega \neq 1$ in the Higgs regime, e.g., the dense QCD.
Since the fundamental matter decouples in the deep confining limit, it would be plausible to determine $O_\Omega = 1$ in the confining regime.
This apparent mismatch in the AB phase $O_\Omega$ may cast doubt on the Higgs-confinement continuity.
Based on this observation, Ref.~{\cite{Cherman:2020hbe}} conjectured that a Higgs-confinement transition exists with an order parameter $O_\Omega$.

\begin{figure}[t]
  \begin{center}
   \includegraphics[width= \linewidth]{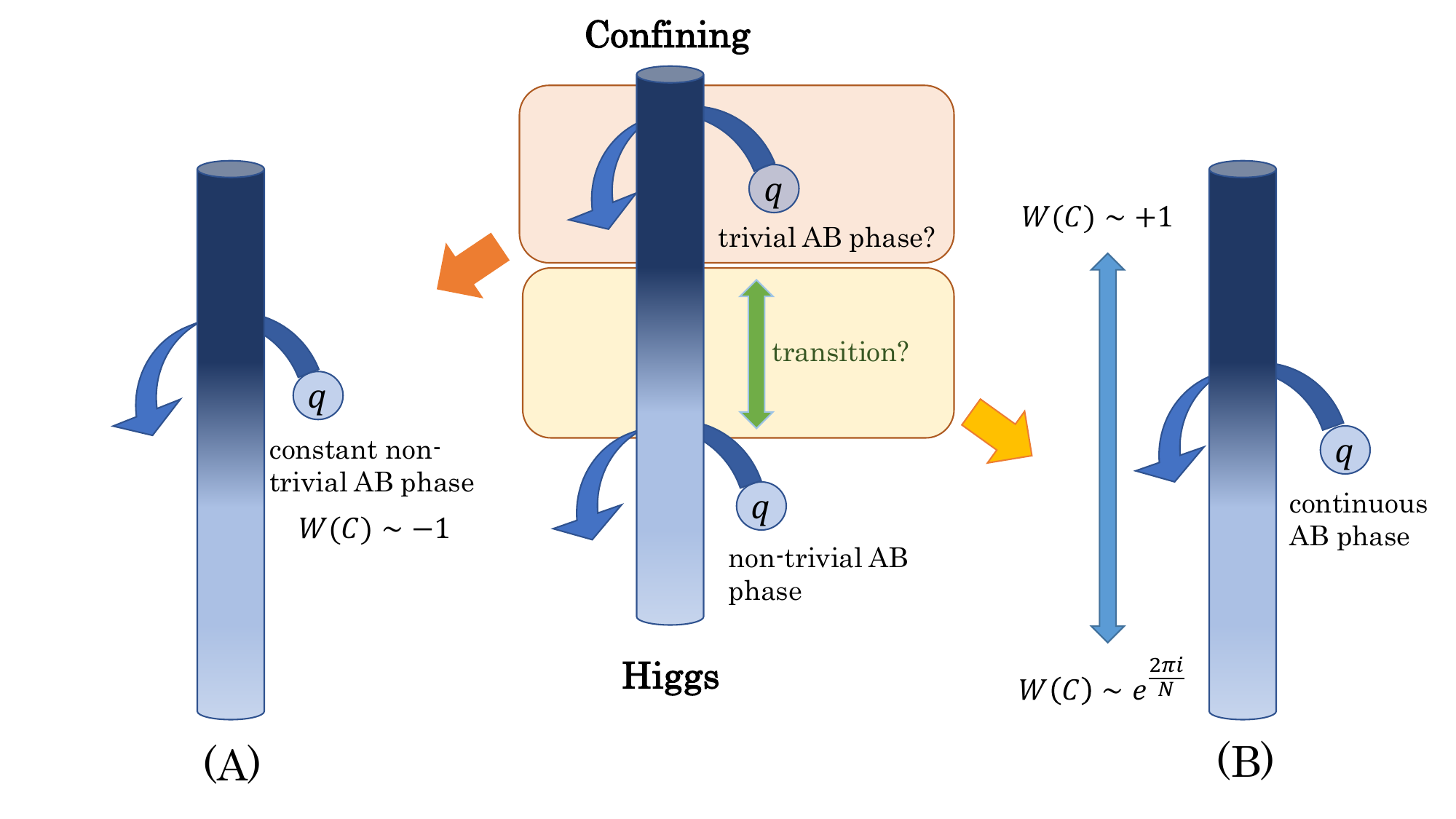}
  \end{center}
   \caption{(center) The apparent discrepancy in the AB phase seems to conflict with the Higgs-confinement continuity.
   Resolving this mismatch, this paper illustrates how the AB phase realizes the continuity:
   (A) If symmetry constrains the AB phase, the AB phase is nontrivial in both Higgs and confining regimes.
   (B) If not, the AB phase continuously changes from $W(C) \sim +1$ (confining) to $W(C) \sim e^{2\pi i/N}$ (Higgs).
   }
    \label{fig:vortex-matching.pdf}
\end{figure}

Here, supporting the Higgs-confinement continuity, we propose the following solution to the mismatch problem (Fig.~\ref{fig:vortex-matching.pdf}):
\begin{enumerate}[label=(\Alph*)]
    \item If there exists a symmetry constraining the AB phase to discrete values, the AB phase is nontrivial even in the confining regime.
    \item If not, the AB phase can continuously interpolate the confining and Higgs limits.
\end{enumerate}
We explicitly demonstrate these claims in the following two examples: (1) a lattice version of the toy model proposed in \cite{Cherman:2020hbe} and (2) an $SU(N)$ lattice gauge model with a fundamental $U(N)$-valued scalar, which is analogous to the gauged Ginzburg-Landau effective model for the CFL phase at $N=3$.
To this end, we calculate the AB phase in two calculable limits: strong coupling and deep Higgs limits, which connect the Higgs and confining regimes.
We find that Claim (A) applies to (1) and (2) at $N=2$ and that Claim (B) applies to (2) for $N>2$.

\section{Nontrivial Aharonov-Bohm phase in the confining regime}

The first example is a lattice version of the toy model proposed in \cite{Cherman:2020hbe}.
This model consists of a $U(1)$ gauge field, charge-$(+1)$ scalar, charge-$(-1)$ scalar, and neutral scalar. The $U(1)$ gauge field is represented by the $U(1)$-valued link variable $U_\ell$. The scalar fields are represented by the $U(1)$-valued fields $\phi_{+, v}$, $\phi_{-, v}$ and $\phi_{0,v}$ defined on vertices.
Its lattice action reads,
    \begin{align}
        S &= \beta \sum_{p:\mathrm{plaquettes}} U_p + \kappa \sum_{\ell:\mathrm{links}} \phi^{*}_{+,x} U_{\ell} \phi_{+,x'} \notag \\
        & + \kappa \sum_{\ell:\mathrm{links}} \phi^{*}_{-,x} U_{\ell}^* \phi_{-,x'} + \kappa_0 \sum_{\ell:\mathrm{links}} \phi^{*}_{0,x}  \phi_{0,x'} \notag \\
        &+ \varepsilon \sum_{v:\mathrm{vertices}} \phi_{+,v} \phi_{-,v} \phi_{0,v}+ \mathrm{c.c.},
    \end{align}
    where $U_p$ expresses the product of link variables $U_\ell$ along the boundary of plaquette $p$, the initial and final points of link $\ell$ are denoted by $x$ and $x'$ respectively, and $\mathrm{c.c.}$ means the complex conjugation.

Note that dynamical vortices of $\phi_+$ and $\phi_-$ are automatically included in the lattice description even if the scalars are $U(1)$-valued, so this lattice model has a confining regime as in the original Fradkin-Shenker model \cite{Fradkin:1978dva}.
The confining regime can be described by small $\kappa$, and the Higgs regime can be described by large $\kappa$.

This model has the following $(U(1)_{gauge} \times U(1)_G)/ \mathbb{Z}_2$ symmetry:
\begin{align}
&U(1)_{gauge}:(\phi_+, \phi_-, \phi_0) \rightarrow (e^{i\alpha} \phi_+, e^{-i\alpha} \phi_-, \phi_0) \notag \\
&U(1)_G:(\phi_+, \phi_-, \phi_0) \rightarrow (e^{i\alpha} \phi_+, e^{i\alpha} \phi_-, e^{-2i\alpha}\phi_0), \label{eq:sym-toymodel}
\end{align}
This $U(1)_G$ symmetry is spontaneously broken by taking large $\kappa_0$ for $d \geq 3$, and its minimal vortex is the $\phi_0$ vortex.
Then, we have $U(1)_G$-broken confining regime for small $\kappa$ and $U(1)_G$-broken Higgs regime for large $\kappa$.
In addition, there exists $\mathbb{Z}_2$ flavor permutation symmetry acting as $\phi_\pm \rightarrow \phi_\mp$ and $U_\ell \rightarrow U_\ell^*$, which constrains the AB phase: $O_\Omega \in \mathbb{Z}_2$.

In what follows, to illustrate Claim (A) in this model, we shall prove that
\begin{align}
    O_\Omega = 
    \begin{cases}
        -1  ~~(\mathrm{deep~Higgs~}\kappa \gg 1) \\
        -1 ~~(\mathrm{strong~coupling~}\beta \ll 1).
    \end{cases}
   \label{eq:main-claim-lattice}
\end{align}
The nontrivial phase in confining regime is new and resolves the AB phase mismatch puzzle of this model.

For later convenience, let us define
\begin{align}
    u_\ell(\phi_{\pm}) := \phi^{*}_{\pm,x} \phi_{\pm,x'} \in U(1), 
\end{align}
which corresponds to the pure gauge link variable.

We first show $O_\Omega = -1$ in the deep Higgs limit.
In this limit, the $U_\ell$ is fixed to maximize the action proportional to $\kappa$:
\begin{align}
    S \supset \kappa \sum_\ell U_\ell \left( u_\ell(\phi_+)   +   u_\ell^*(\phi_-) \right) + \mathrm{c.c.}. \label{eq:deep-higgs-lattice}
\end{align}
We can extract its phase as
\begin{align}
    u_\ell(\phi_+)   +   u_\ell^*(\phi_-) &= C e^{i \varphi_\ell}, \notag \\
    C:= |u_\ell(\phi_+)   +   u_\ell^*(\phi_-)|&,~~e^{i \varphi_\ell} := u_\ell(\phi_+) \sqrt{u_\ell^*(\phi_+ \phi_-)}, \label{eq:polar-form-u}
\end{align}
where we take the principal branch $\operatorname{Re} \sqrt{u_\ell^*(\phi_+ \phi_-)} \geq 0$, which appears from $1+ u_\ell^*(\phi_+ \phi_-) = \sqrt{u_\ell^*(\phi_+ \phi_-)} |1+ u_\ell^*(\phi_+ \phi_-)|$.
Thus, the value of $U_\ell$ maximizing (\ref{eq:deep-higgs-lattice}) is 
\begin{align}
    U_\ell = e^{-i \varphi_\ell} = u_\ell^*(\phi_+) \sqrt{u_\ell(\phi_+ \phi_-)}.
\end{align}
The Wilson loop is then pinned to
\begin{align}
    \prod_{\ell \in \Gamma} U_\ell = \prod_{\ell \in \Gamma} \sqrt{u_\ell(\phi_+ \phi_-)}, \label{eq:deep-higgs-Wilson-loop}
\end{align}
where $\Gamma$ is a closed loop on the lattice, and we have used $\prod_{\ell \in \Gamma} u_\ell(\phi_+) = 1$.

\
To evaluate the AB phase $O_\Omega$ ({\ref{eq:UV_def}}), it is sufficient to consider the asymptotically large Wilson loop $W(C)$ with the fixed winding number of $\phi_0$ along $C$.
The interaction $\varepsilon \phi_+ \phi_- \phi_0$ fixes the asymptotic behavior of the product $\phi_+ \phi_-$: $\phi_+ \phi_- \phi_0 \rightarrow +1$ as $|x| \rightarrow \infty$ \footnote{One can also verify that the integration over $\phi_+$ and $\phi_-$ gives the phase determined by this replacement, when the variation of $\phi_0$ per one link is sufficiently slow. In addition, one can even take large $\varepsilon$ without changing the essence of the problem.}.
Then, this vortex yields a nontrivial phase $\prod_{\ell \in \Gamma} \sqrt{u_\ell(\phi_+ \phi_-)} = -1$, which proves $O_\Omega = -1$ in the deep Higgs limit, in accordance with the continuum result \cite{Cherman:2020hbe}.

Next, to reach the confining regime, let us evaluate the Wilson loop at the strong coupling limit $\beta \rightarrow +0$.
In this limit, the $\{ U_\ell \}$-dependent part of the action is (\ref{eq:deep-higgs-lattice}).
The functional integration is factorized, and a simple computation yields 
\begin{align}
    \int &\prod_\ell dU_\ell \left(\prod_{\ell \in \Gamma} U_\ell \right) e^{\kappa \left(  \sum_\ell U_\ell \left( u_\ell(\phi_+)   +   u_\ell^*(\phi_-) \right) + \mathrm{c.c.} \right)} \notag \\
    &=  \left(\prod_{\ell \in \Gamma} e^{-i \varphi_\ell} I_1 (2 \kappa C) \right)  \times \prod_{\ell \notin \Gamma} I_0 (2 \kappa C)
\end{align}
where $I_0 (2 \kappa C)$ and $I_1 (2 \kappa C)$ are the modified Bessel function of the first kind, and we have used (\ref{eq:polar-form-u}) in the exponent.
Hence, after the $\{ U_\ell \}$ integration, the Wilson loop has the same phase $\prod_{\ell \in \Gamma} e^{-i \varphi_\ell}$ as the deep Higgs limit (\ref{eq:deep-higgs-Wilson-loop}).
This proves $O_\Omega = -1$ in the strong coupling limit (\ref{eq:main-claim-lattice}).


Note that this result is robust in the strong coupling expansion.
The higher order terms in $\beta$, which attach plaquettes connected to the loop $\Gamma$, make only small deformations of the loop $\Gamma$.
The nontrivial phase arises from the asymptotic winding of $\phi_0$ or $\phi_+ \phi_-$, which is clearly invariant under small deformations of the loop.
Besides, small corrections cannot affect $O_\Omega$ due to the $\mathbb{Z}_2$ flavor permutation symmetry.

Finally, let us remark that the $\mathbb{Z}_2$  flavor permutation symmetry is vital to fix the AB phase to $O_\Omega = - 1$.
In the Higgs regime, the $\mathbb{Z}_2$ phase appears because $\phi_+$ and $\phi_-$ equally contribute to the Wilson loop in the above derivation.
Similarly, in the deep confining regime ($\beta \rightarrow +0,~\kappa \ll 1$), each link gives
\begin{align}
    \int& dU_\ell~U_\ell  e^{\kappa \left( U_\ell \left( u_\ell(\phi_+)   +   u_\ell^*(\phi_-) \right) + \mathrm{c.c.} \right)} \notag \\
    &=  \kappa  \left( u_\ell^*(\phi_+)   +   u_\ell(\phi_-) \right) + O(\kappa^2),
\end{align}
which leads to $O_\Omega = -1$ due to the same mechanism as the Higgs regime.
The balance between $\phi_+$ and $\phi_-$ is essential here, and it is guaranteed by the $\mathbb{Z}_2$  flavor permutation symmetry.
By breaking this $\mathbb{Z}_2$ symmetry, e.g., by choosing different weights of the kinetic terms for $\phi_+$ and $\phi_-$, one can change the AB phase freely.

\section{Yang-Mills-Higgs model}

In the second example, we consider an $SU(N)$ Yang-Mills-Higgs model with superfluidity. 
This model consists of $SU(N)$-valued link variables $U_\ell$, a $U(N)$-valued fundamental scalar $\phi_v$, and a $U(1)$-valued gauge-neutral scalar $\phi_{0,v}$ with the action
\begin{align}
    S &= \beta \sum_{p} \operatorname{tr} U_p + \kappa \sum_{\ell} \operatorname{tr} \phi_{x}^\dagger U_\ell \phi_{x'} + \varepsilon \sum_v \phi_{0,v} ^{*} \operatorname{det} \phi_v \notag \\
    &~~~~~~ +\kappa_0 \sum_{\ell} \phi^{*}_{0,x}  \phi_{0,x'} + \mathrm{c.c.}.
\end{align}
This model has the $(SU(N)_{gauge}\times U(N)_f)/\mathbb{Z}_N$ symmetry, acting on $\phi$ and $\phi_0$ as,
\begin{align}
    \phi \rightarrow C \phi F,~~\phi_0 \rightarrow (\operatorname{det}F) \phi_0,
\end{align}
for $C \in SU(N)_{gauge}$ and $F \in U(N)_f$.
The global $U(1) \subset U(N)_f/\mathbb{Z}_N$ symmetry can be spontaneously broken by taking large $\kappa_0$, and we then have a superfluid vortex in both confining and Higgs regimes.

This model at $N=3$ would capture an essential part of the quark-hadron continuity problem.
In the CFL phase, the diquark condensate $\Phi$ is a $3 \times 3$ matrix-valued scalar field, and the symmetries acts on $\Phi$ as $\Phi \rightarrow e^{2i\alpha} F \Phi C^\dagger$, where $C \in SU(3)_{gauge}$ is the color symmetry, $F \in SU(3)_{f}$ is the flavor symmetry, and $e^{i\alpha} \in U(1)$ is the quark-number symmetry. 
The $U(1)$ symmetry is spontaneously broken in the superfluid phases, and the $SU(3)_f$ symmetry is combined with the color symmetry in the CFL phase.
The relevant global symmetry for $\Phi$ can be written as $U(3)$.
Thus, the $SU(3)$ gauge theory with fundamental $U(3)$-valued scalar will be a simple effective model for the CFL phase, analogous to the gauged Ginzburg-Landau-type effective model.
Since vortices are automatically included on the lattice, this model can also describe confining regime.
For the spontaneous breaking of $U(1)$ symmetry in the confining regime, we add the ``dibaryon'' neutral scalar $\phi_0$.

In the deep Higgs limit $\kappa \rightarrow \infty$, the link variable $U_\ell$ is pinned to maximize the Higgs kinetic term. Since $\phi_{x} \phi_{x'}^\dagger  \in U(N)$, it is given by
\begin{align}
    U_\ell = (\operatorname{det} \phi_{x} \phi_{x'}^\dagger)^{-1/N} \phi_{x} \phi_{x'}^\dagger,
\end{align}
where we take the branch $-\frac{2\pi}{2N} \leq \operatorname{Arg}(\operatorname{det} \phi_{x} \phi_{x'}^\dagger)^{-1/N} < \frac{2\pi}{2N}$, which is taken to be nearest to $+1$ in the $U(1)$ circle.
Therefore, the Wilson loop reads
\begin{align}
   \prod_{\ell \in \Gamma} U_\ell = \prod_{\ell \in \Gamma}(\operatorname{det} \phi_{x} \phi_{x'}^\dagger)^{-1/N}.
\end{align}
As before, since a $\phi_0$ vortex rotates $\operatorname{det} \phi: 0 \rightarrow 2\pi$, the Wilson loop yields the phase $e^{\frac{2 \pi i}{N}}$.
Hence we conclude
\begin{align}
    O_\Omega = e^{\frac{2 \pi i}{N}}  ~~(\mathrm{deep~Higgs~}\kappa \gg 1)
   \label{eq:main-claim-YM-Higgs}
\end{align}
in agreement with the nontrivial braiding phases in the Higgs regime \cite{Cherman:2018jir,Cherman:2020hbe}.

For $N=2$, there exists symmetry constraining the holonomy:
\begin{align}
    \phi \rightarrow i \sigma_y \phi,~~ U_\ell \rightarrow U_\ell^*,
\end{align}
which implies $O_\Omega \in \mathbb{Z}_2$ as in the previous section.
In this case, Claim (A) applies, and one can obtain
\begin{align}
    O_\Omega = 
    \begin{cases}
        -1  ~~(\mathrm{deep~Higgs~}\kappa \gg 1) \\
        -1 ~~(\mathrm{strong~coupling~}\beta \ll 1).
    \end{cases}
   \label{eq:main-claim-SU2}
\end{align}
For a simple demonstration, we here examine the deep confining regime ($\beta \rightarrow +0$, $\kappa \ll 1$). The contribution of one link reads
\begin{align}
    &\int dU_\ell 
 \left(U_\ell \right)_{ij} e^{\kappa \operatorname{tr} \phi_{x}^\dagger U_\ell \phi_{x'} +\mathrm{c.c.}} \notag \\
 &= \frac{\kappa}{2} \left( 1 + \operatorname{det} \phi_{x'} \operatorname{det} \phi_{x}^\dagger \right) (\phi_{x} \phi_{x'}^\dagger)_{ij} + O(\kappa^2).
\end{align}
After the cancellation of $(\phi_{x} \phi_{x'}^\dagger)_{ij}$, the Wilson loop has the phase $\prod_\ell (\operatorname{det} \phi_{x'} \operatorname{det} \phi_{x}^\dagger)^{1/2}$, leading to the AB phase $e^{i \pi} = -1$.
Hence, the AB phase should be constant (\ref{eq:main-claim-SU2}) in the strong coupling region.

 \begin{figure}[t]
  \begin{center}
   \includegraphics[width=0.85 \linewidth]{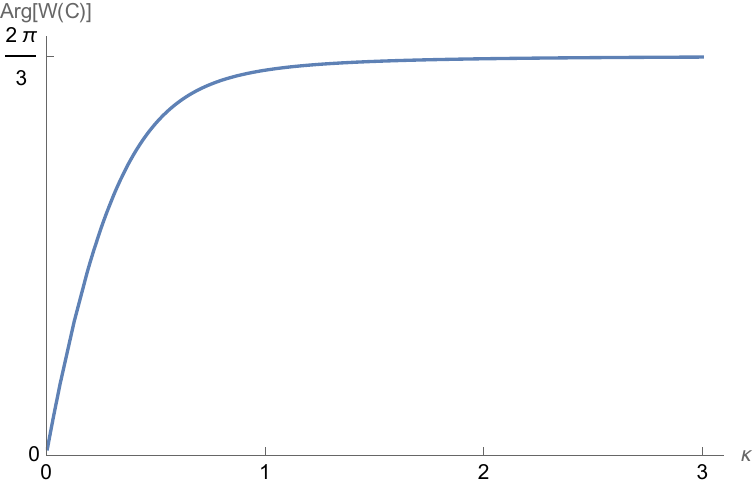}
  \end{center}
   \caption{
   $\kappa$ dependence of the AB phase of the Wilson loop around a vortex at $N=3$, indicating the continuity in the strong coupling limit.
   This plot displays the argument of $\left[ \int_{SU(3)} dU~ U e^{\kappa \operatorname{tr} ( e^{\frac{2 \pi i}{L}} U)+\mathrm{c.c.} }\right]^L$ at $L=100$, which is the AB phase for a vortex changing its phase constantly along the loop of $L$ links.
   }
    \label{fig:AB_phase_w8.pdf}
\end{figure}

For $N>2$, the AB phase at the deep confining limit is trivial $O_\Omega = +1$, but there is no symmetry constraining the value of the AB phase.
In this case, Claim (B) applies: the AB phase can continuously interpolate between $+1$ (confining) and $e^{2\pi i/N}$ (Higgs).
Indeed, at $N=3$, a numerical evaluation of $O_\Omega$ in the strong coupling limit is depicted in Fig.~\ref{fig:AB_phase_w8.pdf}.
Together with (\ref{eq:main-claim-YM-Higgs}), this result indicates the continuity of the AB phase in the deep Higgs ($\kappa \gg 1$) and strong coupling ($\beta \ll 1$) regions, which connect confining and Higgs regimes.


\section{Conclusion and Discussion}



The apparent mismatch of the AB phases has been an open problem for the Higgs-confinement continuity, particularly quark-hadron continuity conjecture.
In general, the AB phase can deviate from the value of a classical solution due to quantum fluctuations unless some symmetry prevents the deviation.
Thus, without such symmetry, the mismatch of AB phases of classical solutions does not contradict the Higgs-confinement continuity from the beginning [Claim (B)].
In the presence of that kind of symmetry, the nontrivial AB phase can also arise in the confining regime, leading to a constant nontrivial AB phase [Claim (A)].
We have explicitly demonstrated this scenario in the two examples: lattice versions of the toy model proposed in Ref.~\cite{Cherman:2020hbe} and the gauged Ginzburg-Landau-type effective model for the CFL phase.
In both cases, the AB phase around the superfluid vortex realizes the Higgs-confinement continuity (Fig.~\ref{fig:vortex-matching.pdf}), which supports that the quark-hadron continuity is still a consistent scenario\footnote{As a related topic, a discrepancy in emergent symmetries also seems to suggest a Higgs-confinement transition.
However, a phase boundary is not necessary unless the emergent symmetries act nontrivially in the infrared limit. 
Thus, the emergent $(d-2)$-form symmetry in the Higgs regime, which acts only on heavy objects (logarithmically confined vortices), e.g., {\cite{Hidaka:2022blq}}, does not distinguish phases {\cite{Hirono:2018fjr}}.}.


Finally, we remark on the occurrence of the nontrivial AB phase in the confining regime.
At first glance, a test particle appears to acquire a trivial AB phase $O_\Omega = 1$ due to the decoupling of charged fields in the low-energy effective theory of the confining regime.
Schematically, the asymptotic behavior of the Wilson loop would consist of the area-law confining term and perimeter-law screening term:
\begin{align}
    \braket{W(C)} \sim (\mathrm{screening~term}) + (\mathrm{confining~term}).
\end{align}
In the confining regime, the area-law confining term is dominant for intermediate-size loops $C$, as described by the effective theory.
However, an asymptotically large Wilson loop is dominated by the screening term, which is beyond the scope of the low-energy effective theory.
The low-energy effective theory by decoupling of charged fields describes local dynamics but does not cover asymptotically large extended objects, which are exactly what the AB phase $O_\Omega$ measures.
The screening term depends on the detailed dynamics of charged fields, and a superfluid vortex can affect it.
Hence, the AB phase can take a nontrivial value even in the confining regime, with a similar mechanism as in the Higgs regime. In this way, the long-distance behavior of the Wilson loop respects the Higgs-confinement continuity.




\section*{Acknowledgements}
The author is greatly indebted to Yuya Tanizaki for fruitful discussions and advice, without which this work could have been neither initiated nor completed.
Y.~H. is supported by JSPS Research Fellowship for Young Scientists Grant No.~20J20215.

\bibliography{references}

\end{document}